\documentstyle[11pt,twoside,colloqOHP2010,epsfig]{article}
\begin{document}
\title{TRAPPIST: a robotic telescope dedicated to the study of planetary systems}
\author{M. Gillon$^{1}$, E. Jehin$^{1}$, P. Magain$^1$, V. Chantry${^1}$, D. Hutsem\'ekers${^1}$, J. Manfroid${^1}$, D. Queloz${^2}$, S. Udry${^2}$}
\affil{$^1$Institut d'Astrophysique et de G\'eophysique,  Universit\'e de Li\`ege,  17 All\'ee du 6 Ao\^ut,  Bat.  B5C, 4000 Li\`ege, Belgium [michael.gillon@ulg.ac.be]}
\affil{$^2$Observatoire de Gen\`eve, Universit\'e de Gen\`eve, 51 Chemin des Maillettes, 1290 Sauverny, Switzerland}
\begin{abstract}
We present here a new robotic telescope called TRAPPIST\footnote{For more details about TRAPPIST, see http//www.astro.ulg.ac.be/sci/Trappist} (TRAnsiting Planets and PlanetesImals Small Telescope). Equipped with a high-quality CCD camera mounted on a 0.6 meter light weight optical tube, TRAPPIST has been installed in April 2010 at the ESO La Silla Observatory (Chile), and is now beginning its scientific program. The science goal of TRAPPIST is the study of planetary systems through two approaches: the detection and study of exoplanets, and the study of comets. We describe here the objectives of the project, the hardware, and we present some of the first results obtained during the commissioning phase.
\end{abstract}
\section{The TRAPPIST project}
The hundreds of exoplanets known today allow us to put our own solar system in the broad context of our galaxy. 
In particular, the subset of known exoplanets that transit their parent stars are key objects  for our understanding of the formation, evolution and properties of planetary systems (e.g., Charbonneau, this volume). 
On the other hand, the objects of our own solar system are and will remain exquisite guides for helping us understand
 the mechanisms of planetary formation and evolution. Notably, comets are most probably remnants of the initial population of planetesimals of the outer part of the protoplanetary disk, and therefore, the study of their physical and chemical properties makes possible a thorough understanding of the conditions that prevailed during the formation of our four giant planets.

Installed last April at the ESO La Silla Observatory (Chile),  our new telescope TRAPPIST  (TRAnsiting Planets and PlanetesImals Small Telescope) is fully dedicated to the study of planetary systems. More specifically, its science goals are the following. 
(1) The photometric search for transits of planets detected by radial velocities (RV). TRAPPIST was designed to be precise enough to detect the transit of a Neptune-size planet in front of a solar-type star, and the transit of a `super-Earth' in front of a red dwarf.
(2) The photometric follow-up of planet candidates found by the transit surveys CoRoT  and WASP.
(3) The characterization of confirmed transiting planets by high-precision eclipse photometry. Transit light curves obtained by TRAPPIST will be used to determine precisely the size of the planet and to possibly detect gravitational perturbations due to an undetected planet. For extremely irradiated giant planets, TRAPPIST will also be used to measure, in the very near-infrared, the planetary emission that is blocked out during the occultation. 
(4) The photometric variability monitoring of HARPS late-type targets to help discriminating astrophysical and planetary RV signals.
(5) The photometric monitoring of bright comets visible from the Southern hemisphere to determine the evolution of their dust and gas production rates as a function of the heliocentric distance. These observations will be performed in narrow-band filters specially designed by NASA for the observing campaign of the great comet Hale-Bopp (Farnham et al. 2000).

TRAPPIST's optical tube unit is a Ritchey-Chretien 0.6 meter telescope with a focal length of 4.8 meter. It is associated with a German equatorial mount that is, thanks to its direct drive system, extremely fast (up to 50 deg/sec), accurate (tracking accuracy without autoguider better than 10" in 10 min), and free of periodic error. 
The instrument is a Peltier-cooled commercial camera equipped with a Fairchild 3041 back-illuminated 2k $\times$ 2k chip. The  pixel scale is 0.64"/pixel. Three read-out modes are available, the shortest read-out time being 2s. The total field of view of the camera is 22' $\times$ 22'. It is associated to a custom-made dual filter wheel. One of the filter wheel contains broad band filters for the exoplanet photometry (Johnson $B$, $V$, $R$, Cousins $Ic$, Sloan $z'$, and a special `$I+z$' filter), while the other contains the narrow-band cometary filters. 

The telescope is protected by a 5 meter diameter dome that was totally refurbished and automatized early 2010. The observatory is fully robotic and equipped with a weather station, a UPS and webcams. 

\section{First results}
 Fig. 1 shows three examples of exoplanet transit light curves obtained by TRAPPIST, for the planets WASP-18b, WASP-19b and WASP-30b (that is in fact a brown dwarf, see Anderson et al. 2010). The obtained photometric precisions per 5 minutes interval are, respectively, 620 ppm for WASP-19 ($I$=11.4), 590 ppm for WASP-30 ($I$=11.4) and 395 ppm for WASP-18 ($I$=8.8).  
 
 As TRAPPIST is totally dedicated to our planetary program, it will be able to reach even better photometric precisions by observing  the eclipses of the same objects as many times as needed, and should be able to meet all its science goals, including the most challenging ones.

\begin{figure}[]
\begin{center}
\epsfig{width=3.85cm,file=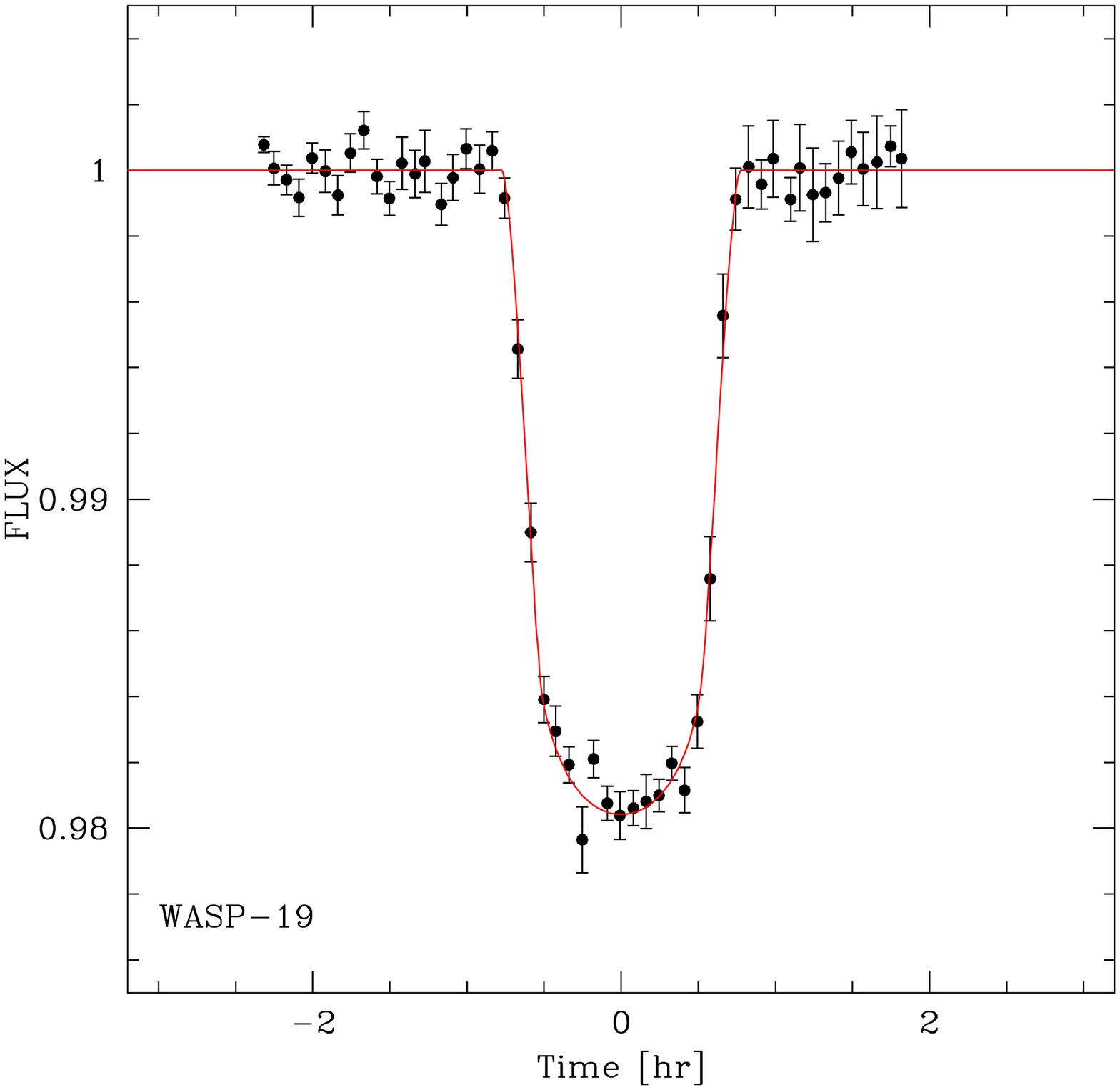}
\epsfig{width=3.85cm,file=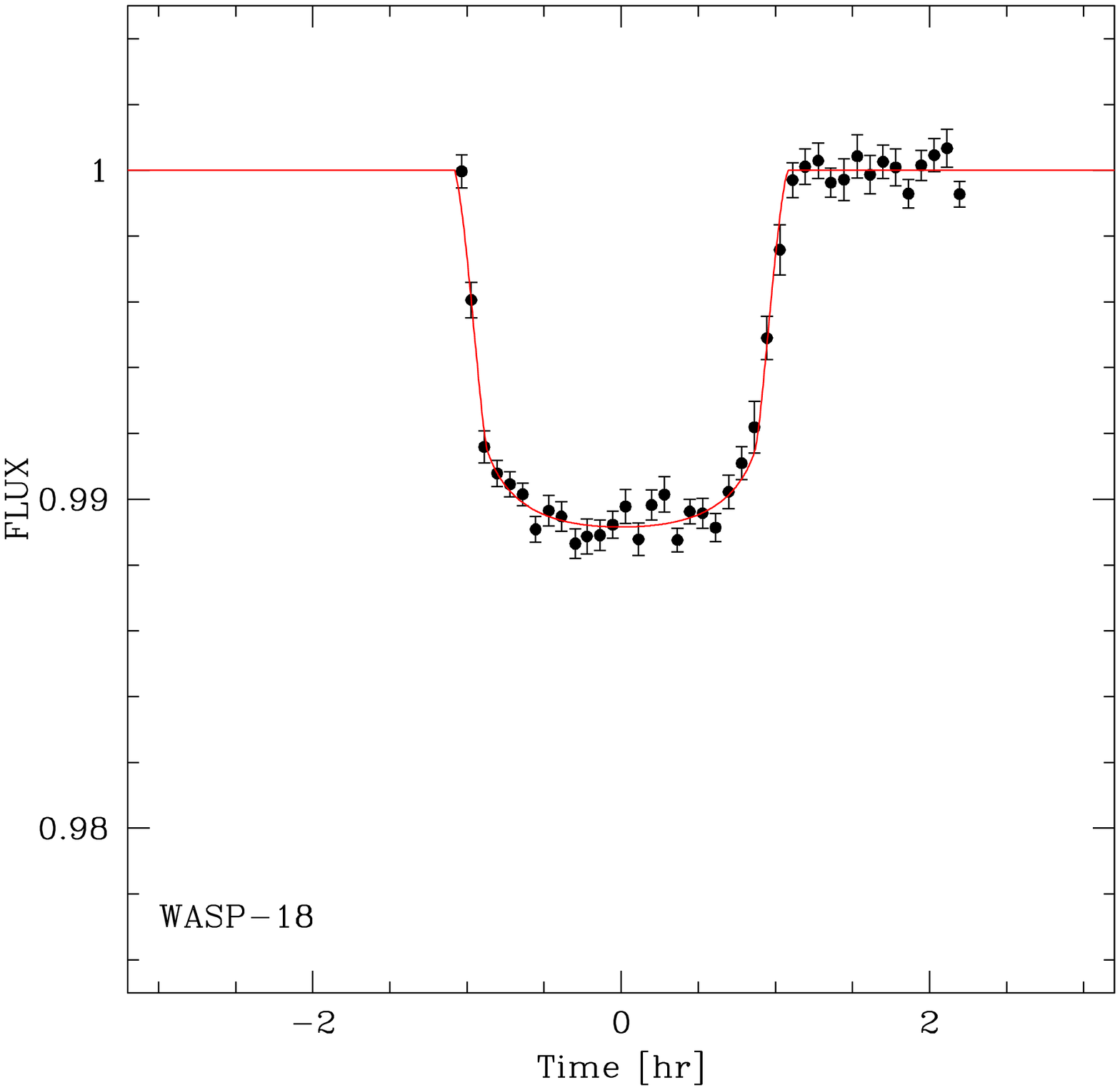}
\epsfig{width=3.85cm,file=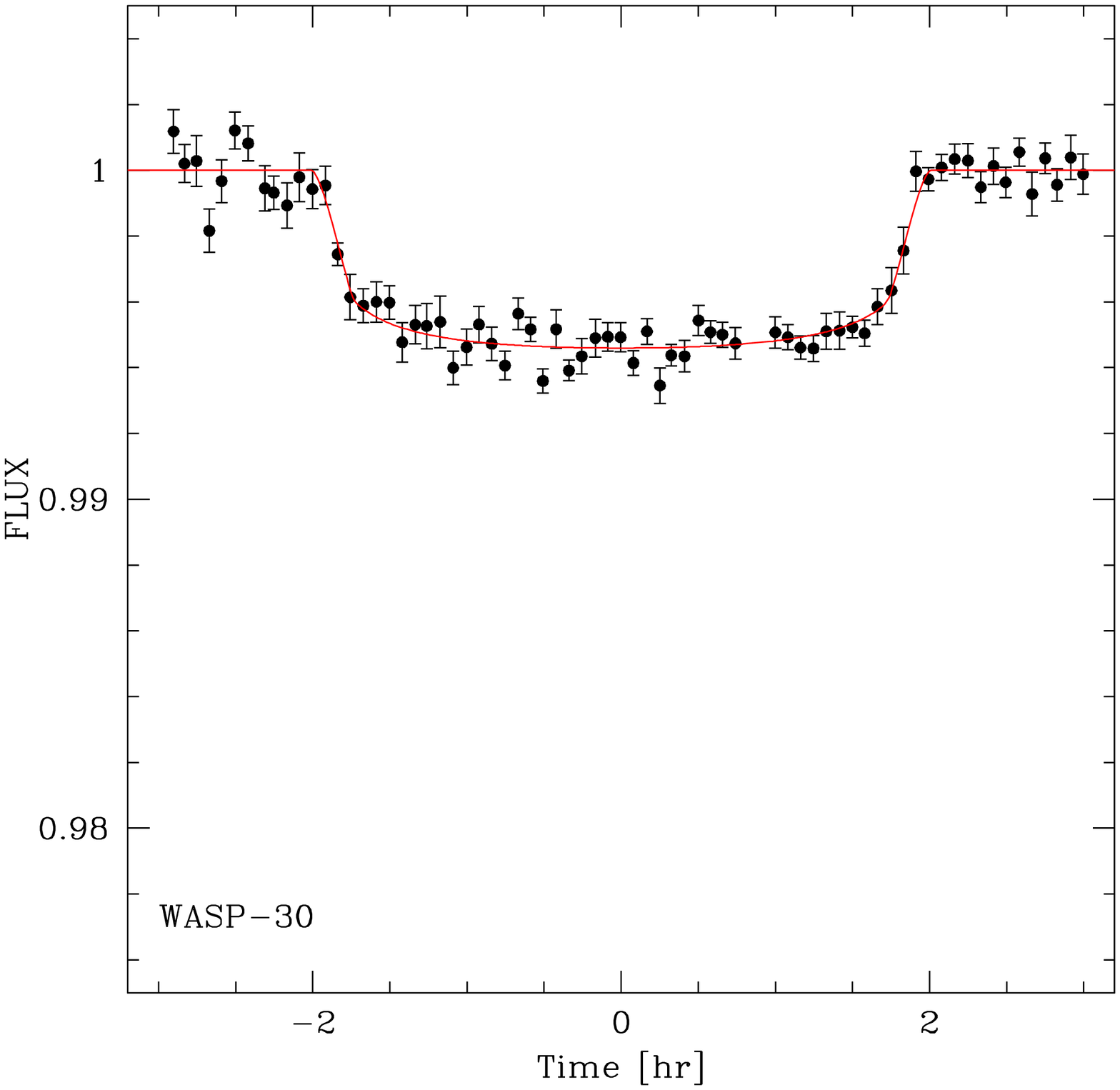}
\caption{From left to right: transit of WASP-18b, WASP-19b and WASP-30b observed by TRAPPIST in the filter `I+z'. All the measurements are binned per time interval of 5 minutes for the sake of clarity. The best-fitting transit models are superimposed in red. }
\end{center}
\end{figure}

\acknowledgments{
M. Gillon and E. Jehin are FNRS Research Associate, D. Hutsem\'ekers is FRNS Senior Research Associate, J. Manfroid is FNRS Research Director, and V. Chantry is FNRS Research Fellow. 
}

\end{document}